\documentclass[a4paper]{spie}  %>>> use this instead for A4 paper
\usepackage{amsmath,amsfonts,amssymb}
\usepackage{graphicx}
\usepackage[colorlinks=true, allcolors=blue]{hyperref}

\title{The Common Path of SOXS (Son of X-Shooter)}

\author[a]{Claudi R.}
\author[b]{Aliverti M.}
\author[a]{Biondi F.}
\author[c]{Munari M.}
\author[c]{Zanmar Sanchez R.}
\author[b]{Campana S.}
\author[d]{Schipani P.}
\author[a]{Baruffolo A.}
\author[e]{Ben--Ami S.}
\author[g,h]{Brucalassi A.}
\author[d]{Capasso G.}
\author[i,c]{Cosentino R.}
\author[j]{D'Alessio F.}
\author[b]{D'Avanzo P.}
\author[f]{Hershko O.}
\author[k]{Kuncarayakti H.}
\author[f]{Rubin A.}
\author[c]{Scuderi S.}
\author[j]{Vitali F.}
\author[m]{Achr\'en J.}
\author[h]{Araiza--Duran J.A.}
\author[n]{Arcavi I.}
\author[b]{Bianco A.}
\author[a]{Cappellaro E.}
\author[d]{Colapietro M.}
\author[d]{Della Valle M.}
\author[f]{Diner O.}
\author[d]{D'Orsi S.}
\author[a]{Fantinel D.}
\author[o]{Fynbo J.}
\author[f]{Gal--Yam A.}
\author[b]{Genoni M.}
\author[p]{Hirvonen M.}
\author[k,l]{Kotilòainen J.}
\author[l]{Kumar T.}
\author[b]{Landoni M.}
\author[p]{Lehti J.}
\author[j]{Li Causi G.}
\author[a]{Marafatto L.}
\author[l]{Matila S.}
\author[b]{Pariani G.}
\author[h]{Pignata G.}
\author[f]{Rappaport M.}
\author[a]{Ricci D.}
\author[b]{Riva M.}
\author[a]{Salasnich B.}
\author[q]{Smartt S.}
\author[a]{Turatto M.}

\affil[a]{INAF--Osservatorio Astronomico di Padova, vicolo Osservatorio 5, 35122 Padova, Italy}
\affil[b]{INAF-- Osservatorio Astronomico di Brera, via Bianchi 46, 23807 Merate (LC), Italy}
\affil[c]{INAF--Osservatorio Astrofisico di Catania, via di Santa Sofia 78, 95123 Catania, Italy}
\affil[d]{INAF--Osservatorio Astronomico di Capodimonte, Salita Moiariello 16, 80131 Napoli, Italy}
\affil[e]{Harvard Smithsonian Center for Astrophysics, Cambridge, USA}
\affil[f]{Weizmann Institute of Science, Herzl St 234, Rehovot, 7610001, Israel}
\affil[g]{ESO, Karl Schwarzschild Strasse 2, D--85748, Garching bei M\"unchen, Germany}
\affil[h]{Universidad Andres Bello, Avda. Republica 252, Santiago, Chile}
\affil[i]{FGG--INAF, TNG, Rambla J.A. Fern\'andez P\'erez 7, 38712 Bre\~na Baja (TF), Spain}
\affil[j]{INAF--Osservatorio Astronomico di Roma, Via Frascati 33, 00078, Monte Porzio Catone (Roma), Italy}
\affil[k]{Finnish Centre for Astronomy with ESO (FINCA), 20014 University of Turku, Finland}
\affil[l]{Tuorla Observatory, Dep. of Physics and Astronomy, 20014 University of Turku, Finland}
\affil[m]{Incident Angle Oy, Capsiankatu 4 A 29, 20320 Turku, Finland}
\affil[n]{Tel Aviv University, Department of Astrophysics, 69978 Tel Aviv, Israel}
\affil[o]{Dark Cosmology Center, Juliane Maries Vej 30, 2100, Copenhagen, Denmark}
\affil[p]{ASRO (Aboa Space Research Oy), Tierankatu 4B, 20520 Turku, Finland}
\affil[q]{Astrophysics Research Centre, Queen's University, Belfast, County Antrim, BT7 1NN, UK}

\authorinfo{Further author information: (Send correspondence to R. Claudi)\\R. Claudi: E-mail: riccardo.claudi@inaf.it}

\begin{document} 
\maketitle

\begin{abstract}
Son of X-Shooter (SOXS) will be a high-efficiency spectrograph with a mean Resolution-Slit product of $\sim 4500$ (goal 5000) over the entire band capable of simultaneously observing the complete spectral range 350-2000 nm. It consists of three scientific arms (the UV-VIS Spectrograph, the NIR Spectrograph and the Acquisition Camera) connected by the Common Path system to the NTT and the Calibration Unit.
The Common Path is the backbone of the instrument and the interface to the NTT Nasmyth focus flange. The light coming from the focus of the telescope is split by the common path optics into the two different optical paths in order to feed the two spectrographs and the acquisition camera.
The instrument project went through the Preliminary Design Review in 2017
and is currently in Final Design Phase (with FDR in July 2018). This paper
outlines the status of the Common Path system and is accompanied by a series of
contributions describing the SOXS design and properties after the
instrument Preliminary Design Review. 
\end{abstract}

\keywords{Spectrograph, Transients, Astronomical Instrumentation, VIS, NIR}

\section{INTRODUCTION}
\label{sec:intro}  % \label{} allows reference to this section
The research on transients has expanded significantly in the past two decades, leading to some of the most recognized discoveries in astrophysics (e.g. gravitational wave events, gamma-ray bursts, super-luminous supernovae, accelerating universe). Nevertheless, so far most of the transient discoveries still lack an adequate spectroscopic follow-up. Thus, it is generally acknowledged that with the availability of so many transient imaging surveys in the next future, the scientific bottleneck is the spectroscopic follow-up observations of transients. Within this context, SOXS aims to significantly contribute bridging this gap. It will be one of the few spectrographs on a dedicated telescope with a significant amount of observing time to characterize astrophysical transients. It is based on the concept of X-Shooter \cite{vernetetal2011} at the VLT but, unlike its “father”, the SOXS science case is heavily focused on transient events. Foremost, it will contribute to the classifications of transients, i.e. supernovae, electromagnetic counterparts of gravitational wave events, neutrino events, tidal disruptions of stars in the gravitational field of supermassive black holes, gamma-ray bursts and fast radio bursts, X-ray binaries and novae, magnetars, but also asteroids and comets, activity in young stellar objects, and blazars and AGN.

\begin{figure} [ht]
   \begin{center}
   \begin{tabular}{c} %% tabular useful for creating an array of images 
   \includegraphics[height=10cm]{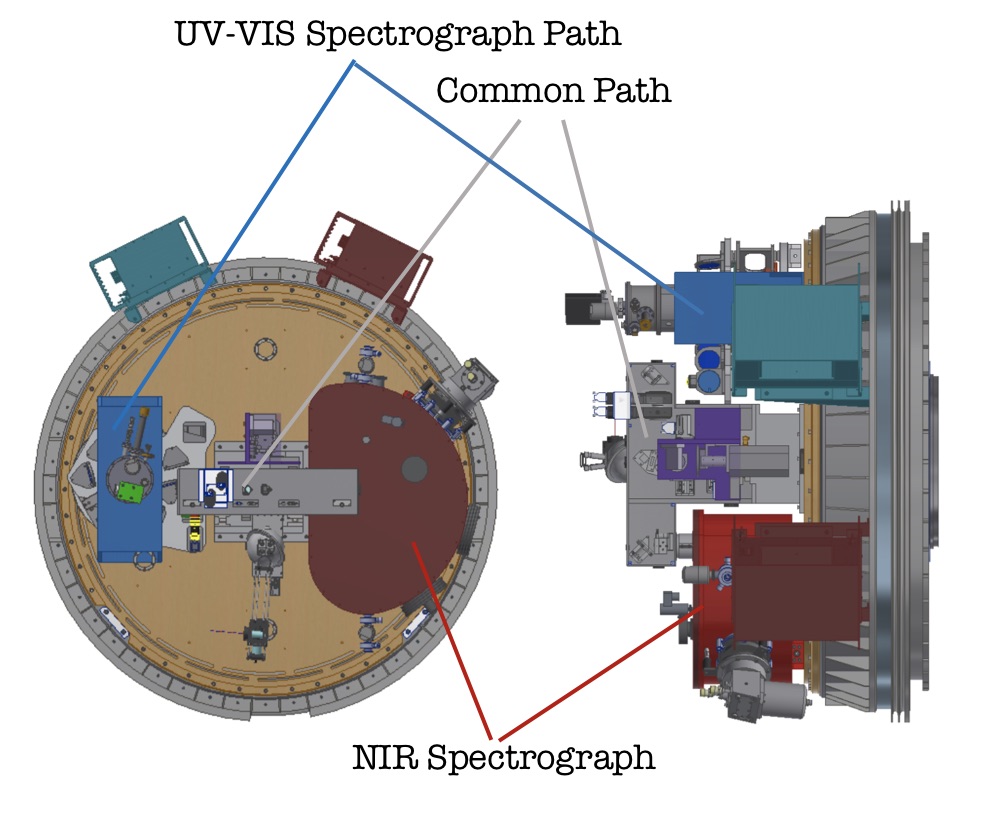}
   \end{tabular}
   \end{center}
   \caption[example] 
%>>>> use \label inside caption to get Fig. number with \ref{}
   { \label{fig:soxs1} 
SOXS: front and side view of the instrument with the identification of the two spectrographs and the common path.}
   \end{figure}

SOXS\cite{schipanietal2018} will simultaneously cover the electromagnetic spectrum from 0.35 to 2.0\ $\mu$m using two arms (UV--VIS and NIR) with a product slit--resolution of $\sim 4500$. The throughput should enable to reach a S/N$\sim 10$ in a 1-hour exposure of an R=20 mag point source. SOXS, that will see its first light at the end of 2020, will be mounted at the Nasmyth focus of NTT replacing SOFI. The whole system (see Figure\ \ref{fig:soxs1}) is constituted by the three main scientific arms: the UV--VIS spectrograph\cite{rubinetal2018, cosentinoetal2018}, the NIR Spectrograph\cite{vitalietal2018} and the acquisition camera\cite{brucalassietal2018}. The three main arms, the calibration box\cite{kuncarayaktietal2018} and the NTT are connected together\cite{biondietal2018} by the Common Path (CP). 

% Note: If compiling with LaTeX+dvipdf, please ensure images generated from 
% other software packages have their bounding boxes set correctly.
The main characteristics of the three scientific arms are listed in Table\ \ref{tab:subsy}.

\begin{table}[ht]
\caption{Main characteristics of the SOXS sub systems connected to the SOXS common path.} 
\label{tab:subsy}
\begin{center}       
\begin{tabular}{|l|c|c|c|} 
\hline
\rule[-1ex]{0pt}{3.5ex}        & AG Camera & UV--VIS& NIR  \\
\hline
\rule[-1ex]{0pt}{3.5ex}  F/\# & 3.6 & 6.5 & 6.5   \\
\hline
\rule[-1ex]{0pt}{3.5ex}  Spectral Range& ugrizY $+$ V& 350--850 nm& 800 -- 2000 nm  \\
\hline
\rule[-1ex]{0pt}{3.5ex}  Resolution& $-$ & $3500 - 7000$ & 5000  \\
\hline
\rule[-1ex]{0pt}{3.5ex}  Pixel Scale (arcsec/px)& 0.205 & 0.139 & 0.164  \\
\hline 
\rule[-1ex]{0pt}{3.5ex}  Detector & Andor& e2v CCD44--82 2k$\times$ 4k& H2RG\ 2k $\times$ 2k  \\
\hline 
\rule[-1ex]{0pt}{3.5ex}  Pixel Size ($\mu$m) & 13.0& 15.0& H2RG\ 18.0  \\
\hline 
\end{tabular}
\end{center}
\end{table}

\section{Common Path}
\label{sec:cp}  % \label{} allows reference to this section
\begin{figure} [ht]
   \begin{center}
   \begin{tabular}{c} %% tabular useful for creating an array of images 
   \includegraphics[height=10cm]{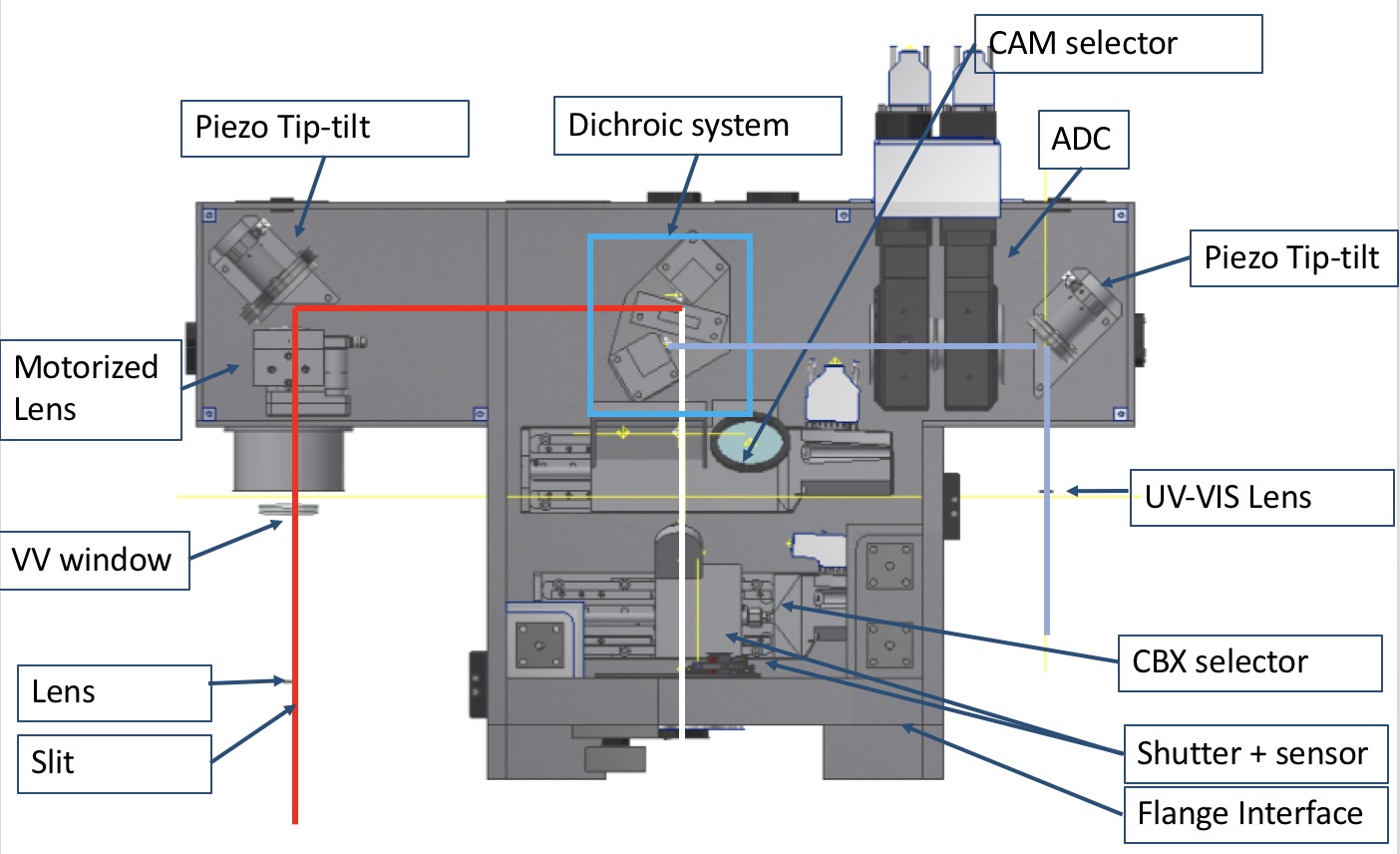}
   \end{tabular}
   \end{center}
   \caption[example] 
%>>>> use \label inside caption to get Fig. number with \ref{}
   { \label{fig:soxs2} 
The sketch of the Common Path sub system. In the picture the several components of the sub system are highlighted.}
   \end{figure}

The Common Path relays the light from the NTT Focal Plane to the entrance of the two spectrographs (UV-VIS and NIR). It selects the wavelength range for the spectrographs using a dichroic and changes the focal ratio of the beam coming from the telescope (F/11) to one suitable for both the spectrographs. A sketch of the common path opto--mechanical design is shown in in Figure\ \ref{fig:soxs2} (overall dimensions are about 650x350 mm) while the main CP parameters are described in Table\ \ref{tab:cpch}. 

% Note: If compiling with LaTeX+dvipdf, please ensure images generated from 
% other software packages have their bounding boxes set correctly.

After the dichroic, two flat folding mirrors direct the light towards two distinct UV-VIS and NIR arms. In the UV-VIS path, the light coming from the folding mirror goes to the ADC assembly, that, strictly speaking, is not composed only of two double prisms correcting the atmospheric dispersion but also of two doublets glued on the prisms. The two doublets (the first one having an aspherical surface) create a collimated beam for the ADC and transform the telescope F/11 beam into an F/6.5. After the ADC, the beam is reflected by the tip/tilt mirror mounted on a piezo stage. Finally, a field lens matches the exit pupil onto the UV-VIS spectrograph pupil. The adopted glasses assure a good transmission ($>80\ \%$) in the UV-VIS side of the spectrum. For the same reason, the dichroic is used in reflection for this wavelength range, in order to give the largest choice of materials for the substrate.

The near infrared path is very similar. It does not include an ADC as mentioned previously. A doublet reduces the telescope F/11 beam to an F/6.5 beam. The doublet has an aspherical surface deemed feasible for manufacturers. A flat tip-tilt folding mirror based on a piezo stage relays the light towards the slit. A flat window is used at the entrance of the spectrograph dewar, with a cold stop after the window itself to reduce the noise. A field lens, placed near the slit, remaps the telescope pupil on the grating of the spectrograph, as in the UV-VIS arm.

\begin{table}[ht]
\caption{Main characteristics of the SOXS common path.} 
\label{tab:cpch}
\begin{center}       
\begin{tabular}{|l|c|} 
\hline
\rule[-1ex]{0pt}{3.5ex}        Input F/\#& 11  \\
\hline
\rule[-1ex]{0pt}{3.5ex}  Field of View& $12\times12$ arcsec   \\
\hline
\rule[-1ex]{0pt}{3.5ex}  Output F/\#& 6.5  \\
\hline
\rule[-1ex]{0pt}{3.5ex}  Image Scale & $110\ \mu$m/arcsec  \\
\hline
\rule[-1ex]{0pt}{3.5ex}  Wavelength range & 350--850\ nm\ (UV--VIS); 800--2000\ nm\ (NIR)  \\
\hline 
\end{tabular}
\end{center}
\end{table}

\section{CP Optics}
\label{sec:optics}  % \label{} allows reference to this section
\begin{figure} [ht]
   \begin{center}
   \begin{tabular}{c} %% tabular useful for creating an array of images 
   \includegraphics[height=7cm]{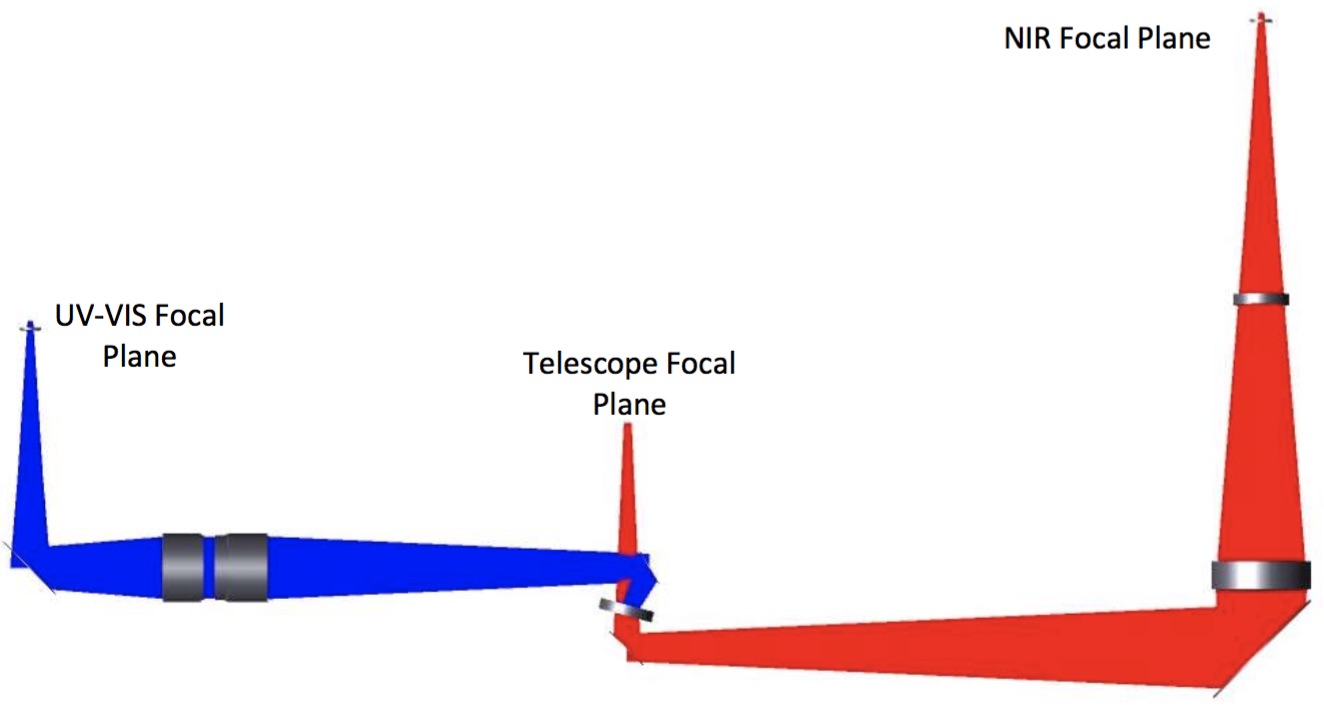}
   \end{tabular}
   \end{center}
   \caption[example] 
%>>>> use \label inside caption to get Fig. number with \ref{}
   { \label{fig:oplay} 
The CP optical layout. The blue path is the optical beam that feeds the UV--VIS spectrograph. The read beam is the optical beam that feeds the NIR spectrograph. The optics are shown without their mechanical holders.}
   \end{figure}

The CP optical layout is shown in Figure\ \ref{fig:oplay} where both the UV--VIS (blue) and NIR (red)  optical beams are indicated. In this Section we briefly describe the optical design of the CP describing the two arms separately; the whole optical design of SOXS is reported in Ref.\ \citenum{zanmarsanchezetal2018}.

% Note: If compiling with LaTeX+dvipdf, please ensure images generated from 
% other software packages have their bounding boxes set correctly.

% Note: If compiling with LaTeX+dvipdf, please ensure images generated from 
% other software packages have their bounding boxes set correctly.
   \begin{figure} [ht]
   \begin{center}
   \begin{tabular}{c} %% tabular useful for creating an array of images 
   \includegraphics[height=6cm]{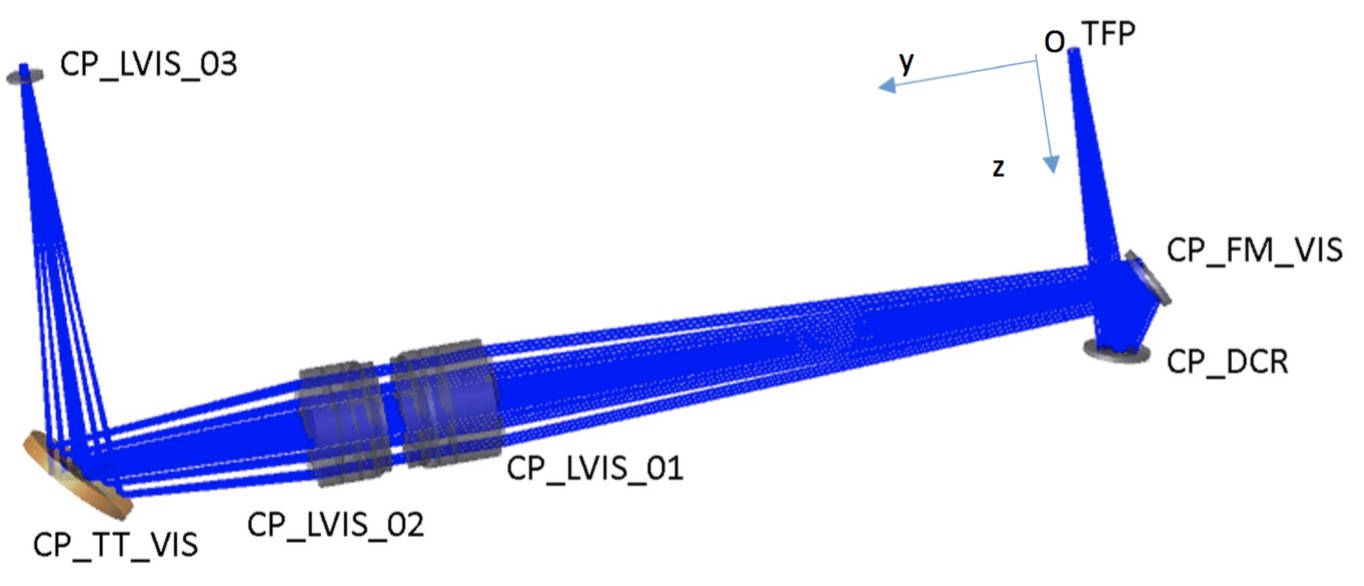}
   \end{tabular}
   \end{center}
   \caption[example] 
%>>>> use \label inside caption to get Fig. number with \ref{}
   { \label{fig:uvisop} 
The optical layout of the UV-VIS feeding arm.}
   \end{figure}

\subsection{The visible arm}
\label{subsec:visarm} 
The CP Visible arm is depicted in Figure\ \ref{fig:uvisop}. It is the optical path towards the UV-VIS spectrograph. The light coming from the Telescope Focal Plane (TFP) is reflected by the CP--Dichroic (CP\_DCR) and a flat folding mirror (CP\_FM\_VIS) into the ADC assembly (CP\_LVIS\_01 and CP\_LVIS\_02). The ADC assembly is actually composed not only of two double prisms (correcting the atmospheric dispersion) but also of two doublets, glued on the prisms (see Section\ \ref{subsec:adc}). The two doublets (the first one having an aspherical surface) create a collimated beam for the ADC and transform the telescope F/11 beam into an F/6.5. The aspherical surface has a small deviation of less than $10\ \mu$m with respect to the Best Fitting Sphere (BSF).
After the ADC, the beam is reflected by the tip/tilt mirror (CP\_TT\_VIS). Finally, a field lens (CP\_LVIS\_03) matches the exit pupil onto the UV-VIS spectrograph pupil.

\begin{figure} [ht]
   \begin{center}
   \begin{tabular}{c} %% tabular useful for creating an array of images 
   \includegraphics[height=8cm]{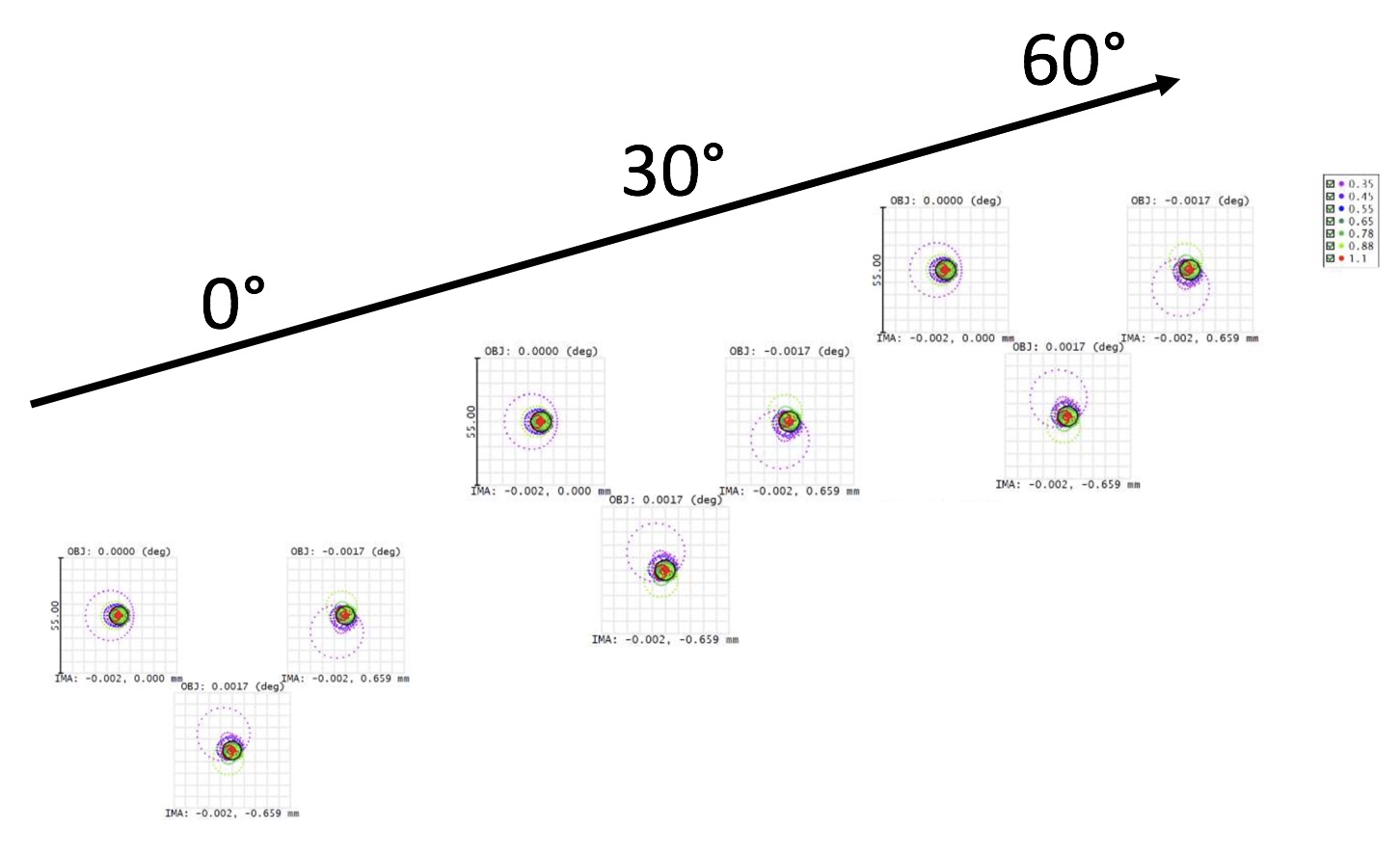}
   \end{tabular}
   \end{center}
   \caption[example] 
%>>>> use \label inside caption to get Fig. number with \ref{}
   { \label{fig:uvopqua} 
The optical quality of the overall visible arm. The spot diagrams are shown for three different zenithal angles.}
   \end{figure}

The optical prescriptions of each optical element of the visible arm are listed in Table\ \ref{tab:cpoppre}. The overall image quality of the visible arm of the CP is shown in Figure\ \ref{fig:uvopqua} considering the spot diagrams achievable at 3 different zenithal angles: $0^\circ$, $30^\circ$ and $60^\circ$ respectively.

\begin{table}[ht]
\caption{Optical prescription of the elements of the CP visisble arm} 
\label{tab:cpoppre}
\begin{center}       
\begin{tabular}{|l|c|c|c|c|} 
\hline
\rule[-1ex]{0pt}{3.5ex}  Element & Curvature Radius& Thickness& Material& Free Aperture  \\
\hline
\rule[-1ex]{0pt}{3.5ex}          &       (mm)& (mm)&       & (mm)  \\
\hline
\rule[-1ex]{0pt}{3.5ex}  TFP     &       &  85.000   &      &        \\
\hline
\rule[-1ex]{0pt}{3.5ex}  CP\_DCR  &  Flat  & $-20.000$& Mirror &16.00\\
\hline
\rule[-1ex]{0pt}{3.5ex}  CP\_FM\_VIS& Flat & 176.290& Mirror & 18.00\\
\hline
\rule[-1ex]{0pt}{3.5ex}  CP\_LVIS\_01 S1& 72.306& 15.000& CAF2 &32.40\\
\hline 
\rule[-1ex]{0pt}{3.5ex}  CP\_LVIS\_01 S2& $-61.880$& 3.000& BAL35Y &32.40\\
\hline 
\rule[-1ex]{0pt}{3.5ex}  CP\_LVIS\_01 S3& Flat& 4.000& BAL15Y &31.60\\
\hline 
\rule[-1ex]{0pt}{3.5ex}  CP\_LVIS\_01 S4& Flat tilt=$-9.369^\circ$& 4.000& S--FPL51Y &31.60\\
\hline
\rule[-1ex]{0pt}{3.5ex}  CP\_LVIS\_01 S5& Flat tilt=$1.146^\circ$& 5.000&        &31.60\\
\hline 
\rule[-1ex]{0pt}{3.5ex}  CP\_LVIS\_02 S1& Flat tilt=$1.146^\circ$& 4.000& S--FPL51Y &31.60\\
\hline
\rule[-1ex]{0pt}{3.5ex}  CP\_LVIS\_02 S2& Flat tilt=$-9.369^\circ$& 4.000& BAL15Y &31.60\\
\hline
\rule[-1ex]{0pt}{3.5ex}  CP\_LVIS\_02 S3& Flat & 3.000& BSM51Y &31.60\\
\hline
\rule[-1ex]{0pt}{3.5ex}  CP\_LVIS\_02 S4& 50.121& 10.000& S--FPL51Y &32.00\\
\hline
\rule[-1ex]{0pt}{3.5ex}  CP\_LVIS\_02 S5& $-63.723$& 60.000&   &32.00\\
\hline
\rule[-1ex]{0pt}{3.5ex}  CP\_TT\_VIS & Flat & $-110.457$& MIRROR &32.00\\
\hline
\rule[-1ex]{0pt}{3.5ex}  CP\_LVIS\_VIS S1 & $-31.500$ & $-1.500$& SILICA &10.00\\
\hline
\rule[-1ex]{0pt}{3.5ex}  CP\_LVIS\_VIS S2&   & $-3.032$&   &10.00\\
\hline
\rule[-1ex]{0pt}{3.5ex}  UV--VIS entrance Slit&   & &   &\\
\hline
\end{tabular}
\end{center}
\end{table}

% Note: If compiling with LaTeX+dvipdf, please ensure images generated from 
% other software packages have their bounding boxes set correctly.

All the spots are enclosed in a square of $55\ \mu$m side corresponding to 0.5 arcsec. In each figure, the three spot diagrams are one for each position in the field: center, bottom and top of the 1 arcsec slit of the UV--VIS spectrograph. Different colors represent different wavelengths. The included wavelength of $1.1\ \mu$m in the spot diagrams is a reference to the NIR arm and it is used to check the residual chromatic error.

\subsubsection{The Atmospheric Dispersion Corrector (ADC)}
\label{subsec:adc} 
\begin{figure} [ht]
   \begin{center}
   \begin{tabular}{c} %% tabular useful for creating an array of images 
   \includegraphics[height=8cm]{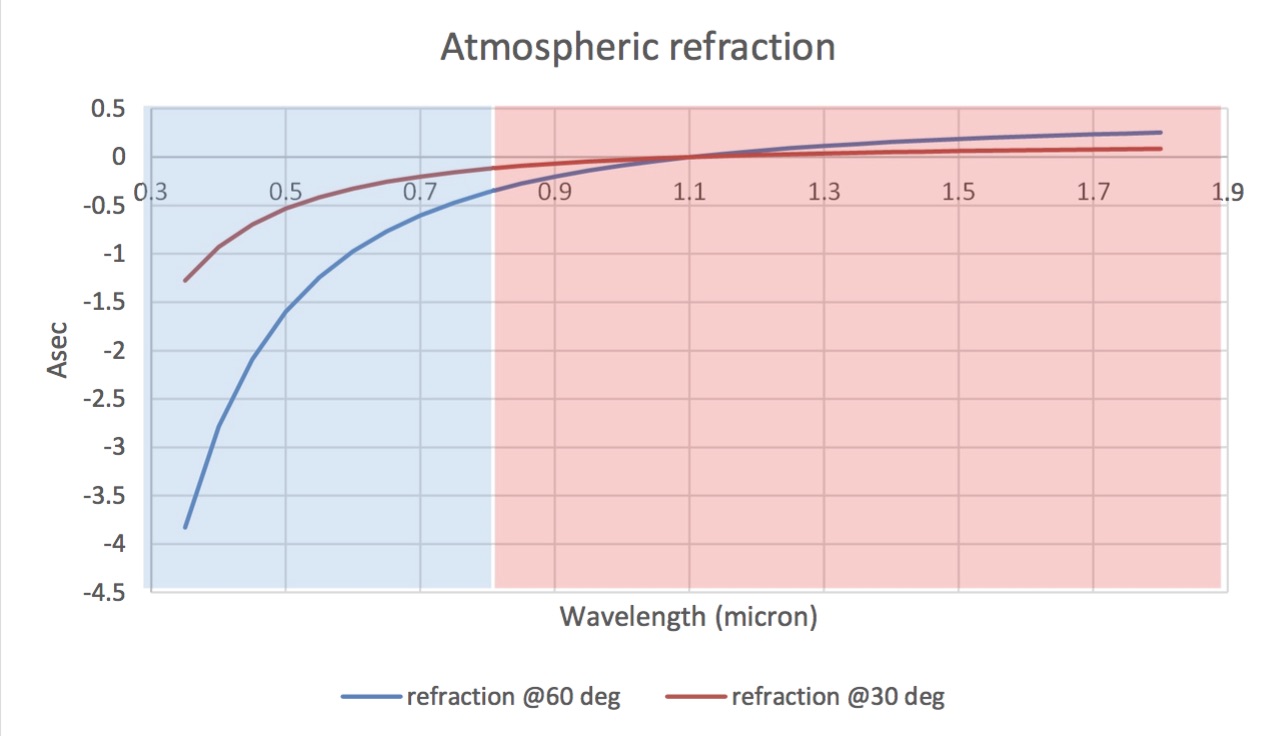}
   \end{tabular}
   \end{center}
   \caption[example] 
%>>>> use \label inside caption to get Fig. number with \ref{}
   { \label{fig:adc} 
The trend of the atmospheric refraction in the visible (blue region) and in the NIR (red region) of the SOXS active wavelength range. The two curves are for different angles.}
   \end{figure}

The CP visible arm, in contrast to the NIR arm, hosts an Atmospheric Dispersion Corrector (ADC). As it is possible to see in Figure\ \ref{fig:adc}, the atmospheric dispersion in the NIR wavelength range (the red shaded region in the Figure) is only $\sim 0.2$\ arcsec at zenith angle z$=30^\circ$ and $\sim 0.5$\ arcsec at z$=60^\circ$. On the contrary, in the visible wavelength range (the blue shaded area in Figure\ \ref{fig:adc}) the same conditions lead to dispersions of ~1.2 and ~3.6 arcsec (values calculated following Ref.\ \citenum{filippenko1982}, using T$=10^\circ$\ C, P$=770$\ mb, H$=44\%$) making mandatory the use of an ADC in the visible range.

% Note: If compiling with LaTeX+dvipdf, please ensure images generated from 
% other software packages have their bounding boxes set correctly.

The SOXS ADC is composed of pair of prisms (each pair composed of two different glasses) counter-rotating to correct the dispersion.
The rotation angle ($\theta$) of the first prism (and the corresponding counter-rotation of the second prism) depends on the zenith angle and the atmospheric conditions and it can be modeled with the following function:

$$\theta = \arccos [(3.6 \times 10^{-1}-1.2 \times 10^{-3}T+ 4.4\times10^{-4}P) \tan(z)] $$

$T$ is the temperature in Kelvin, $P$ the pressure in mbar and $z$ is the zenith angle.
The two doublets of the CP visible arm are glued directly on the prisms. This, besides increasing the transmission of the system and allowing for the elimination of two glass--air interfaces, makes the ADC a simple and compact system. The material description of the ADC (CP\_LVIS\_$\star$) is embedded in the Table\ \ref{tab:cpoppre}.

\subsection{The NIR arm}
\label{subsec:NIR} 
\begin{figure} [ht]
   \begin{center}
   \begin{tabular}{c} %% tabular useful for creating an array of images 
   \includegraphics[height=8cm]{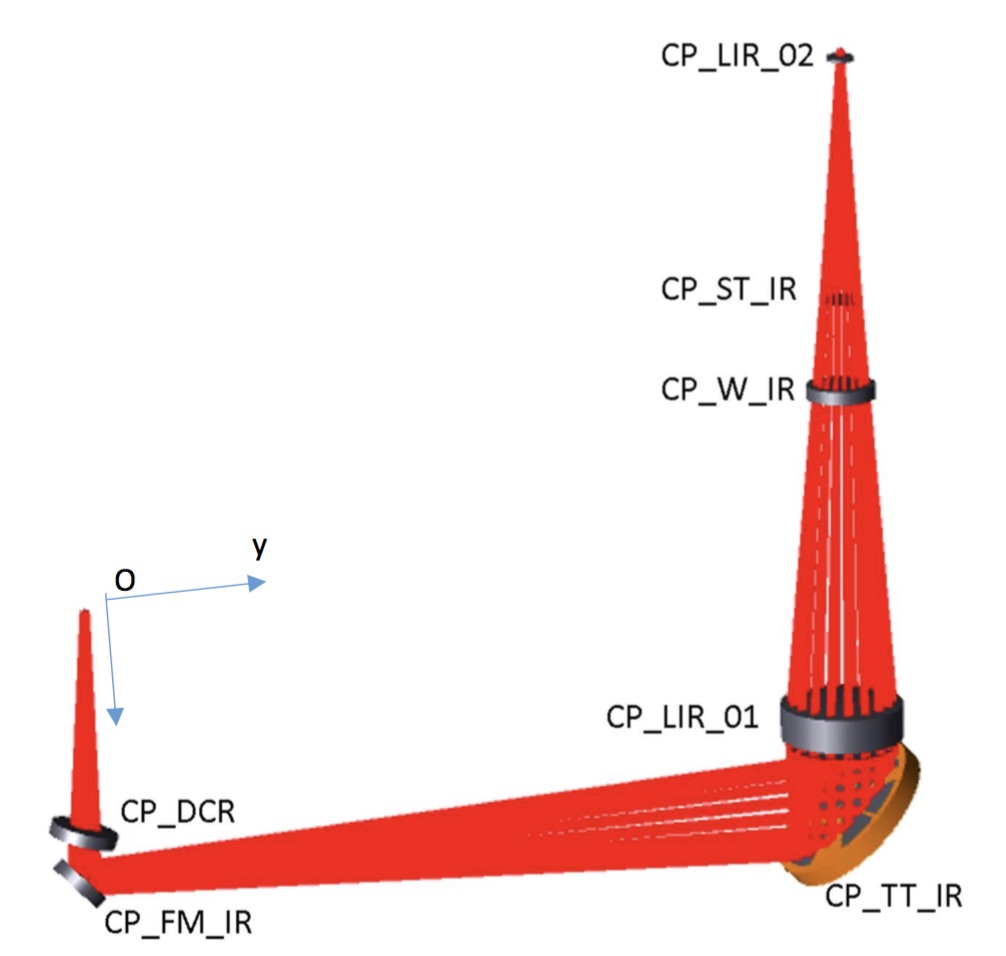}
   \end{tabular}
   \end{center}
   \caption[example] 
%>>>> use \label inside caption to get Fig. number with \ref{}
   { \label{fig:nirarm} The optical scheme of the NIR arm optical path. All the optics are shown without their mechanical holders.}
   \end{figure}

CP-NIR arm is similar to the UV-VIS arm (same first order parameters, except the wavelength range). As discussed in Section\ \ref{subsec:adc}, the NIR arm does not include an ADC, since the atmospheric dispersion is less severe than in the UV--VIS range and is considered acceptable (see Figure \ref{fig:adc}) in the instrument system trade--off. The CP NIR Arm includes a doublet (CP\_LIR\_01) in order to reduce the telescope F/11 beam to an F/6.5 beam. The doublet has an aspherical surface deviating less than $4\ \mu$m with respect to the BFS, deemed feasible to manufacture. Two flat mirrors (CP\_FM\_IR and CP\_TT\_IR) relay the light to the slit. In order to allow the entering of light in the NIR spectrograph dewar, the CP NIR Arm includes a flat window (CP\_W\_IR), and to reduce the noise in the the NIR spectrograph, a cold stop has been introduced after the window itself (CP\_ST\_IR). A field lens (CP\_LIR\_02), placed near the slit, remaps the telescope pupil on the grating of the spectrograph. The whole CP NIR Arm layout is presented in Figure\ \ref{fig:nirarm}.

% Note: If compiling with LaTeX+dvipdf, please ensure images generated from 
% other software packages have their bounding boxes set correctly.

\begin{figure} [ht]
   \begin{center}
   \begin{tabular}{c} %% tabular useful for creating an array of images 
   \includegraphics[height=8cm]{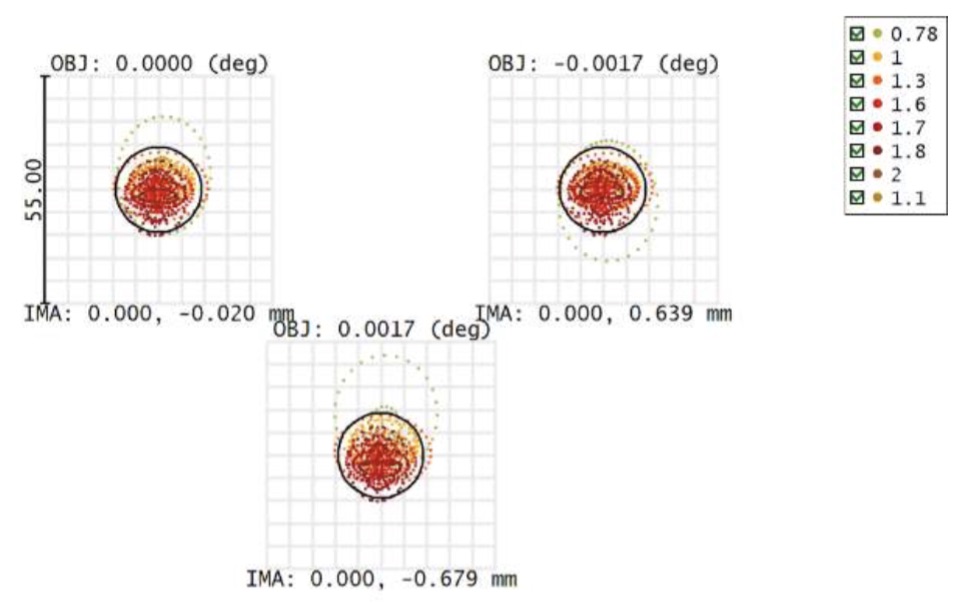}
   \end{tabular}
   \end{center}
   \caption[example] 
%>>>> use \label inside caption to get Fig. number with \ref{}
   { \label{fig:niropqua} The optical quality of the overall visible arm for three different position on the NIR Spectrograph slit.}
   \end{figure}

The optical prescriptions of each optical element of the NIR arm are listed in Table\ \ref{tab:cponir}. The achievable image quality of the overall NIR arm of the CP is shown in Figure\ \ref{fig:niropqua}, by means the spot diagrams  obtained for three positions on the slit (center and two extremes). The boxes of $55\ \mu$m correspond to 0.5 arcsec.

\begin{table}[ht]
\caption{Optical prescription of the elements of the CP visible arm} 
\label{tab:cponir}
\begin{center}       
\begin{tabular}{|l|c|c|c|c|} 
\hline
\rule[-1ex]{0pt}{3.5ex}  Element & Curvature Radius& Thickness& Material& Free Aperture  \\
\hline
\rule[-1ex]{0pt}{3.5ex}          &       (mm)& (mm)&       & (mm)  \\
\hline
\rule[-1ex]{0pt}{3.5ex}  TFP     &       &  85.000   &      &        \\
\hline
\rule[-1ex]{0pt}{3.5ex}  CP\_DCR  &    & 5.000 & SILICA &24.00\\
\hline
\rule[-1ex]{0pt}{3.5ex}  CP\_DCR 2$^{nd}$ Surf&  & 15.000&   & 24.00\\
\hline
\rule[-1ex]{0pt}{3.5ex}  CP\_FM\_IR & & $-295.430$& MIRROR &20.00\\
\hline 
\rule[-1ex]{0pt}{3.5ex}  CP\_TT\_IR &  & 24.000& Mirror &62.00\\
\hline 
\rule[-1ex]{0pt}{3.5ex}  CP\_LIR\_01 S1& 75.515& 5.000& S--TIM2 &45.40\\
\hline 
\rule[-1ex]{0pt}{3.5ex}  CP\_LIR\_01 S2& 46.778& 12.235& CAF2 &45.40\\
\hline 
\rule[-1ex]{0pt}{3.5ex}  CP\_LIR\_01 S3& $-22.844$& 118.367&  &45.40\\
\hline 
\rule[-1ex]{0pt}{3.5ex}  CP\_W\_IR &  & 5.000& SILICA &24.00\\
\hline
\rule[-1ex]{0pt}{3.5ex}  CP\_W\_IR 2$^{nd}$ Surf&    &33.000&        &24.00\\
\hline 
\rule[-1ex]{0pt}{3.5ex}  CP\_ST\_IR &    & 95.572&  &15.50\\
\hline
\rule[-1ex]{0pt}{3.5ex}  CP\_LIR\_02 S1&  41.50& 1.500& SILICA &8.00\\
\hline
\rule[-1ex]{0pt}{3.5ex}  CP\_LIR\_02 S2&  & 3.163&  &8.00\\
\hline
\rule[-1ex]{0pt}{3.5ex}  NIR entrance Slit&   & &   &\\
\hline
\end{tabular}
\end{center}
\end{table}

% Note: If compiling with LaTeX+dvipdf, please ensure images generated from 
% other software packages have their bounding boxes set correctly.

\subsection{Mechanics}
\label{subsec:mech}  % \label{} allows reference to this section
\begin{figure} [ht]
   \begin{center}
   \begin{tabular}{c} %% tabular useful for creating an array of images 
   \includegraphics[height=5cm]{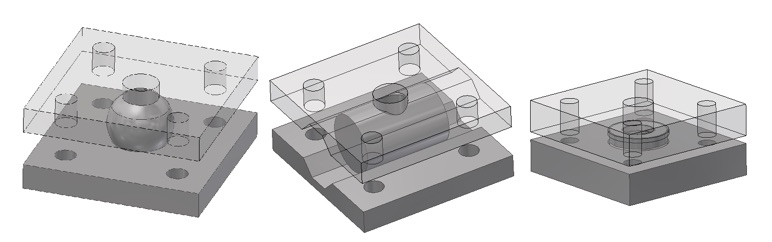}
   \end{tabular}
   \end{center}
   \caption[example] 
%>>>> use \label inside caption to get Fig. number with \ref{}
   { \label{fig:km} Schematic view of the 3 parts of KM mount system. From left to right each element fixes 3, 2 and 1 degree of freedom. A large version is used for CP, NIR and UV-VIS while a smaller one is used for CAM and CBX.}
   \end{figure}

The whole mechanical design of SOXS is described fully in the Ref.\ \citenum{alivertietal2018}. The CP of SOXS is composed of a T--shaped structure made of 6061-T6 aluminum. The structure includes one set of large kinematic mounts (KM) used as interface between CP and interface flange (IF) and two sets of small KM as interface between CP and acquisition Camera and calibration box subsystems. The structure of the KM is shown in Figure\ \ref{fig:km}.

% Note: If compiling with LaTeX+dvipdf, please ensure images generated from 
% other software packages have their bounding boxes set correctly.

\begin{figure} [ht]
   \begin{center}
   \begin{tabular}{c} %% tabular useful for creating an array of images 
   \includegraphics[height=8cm]{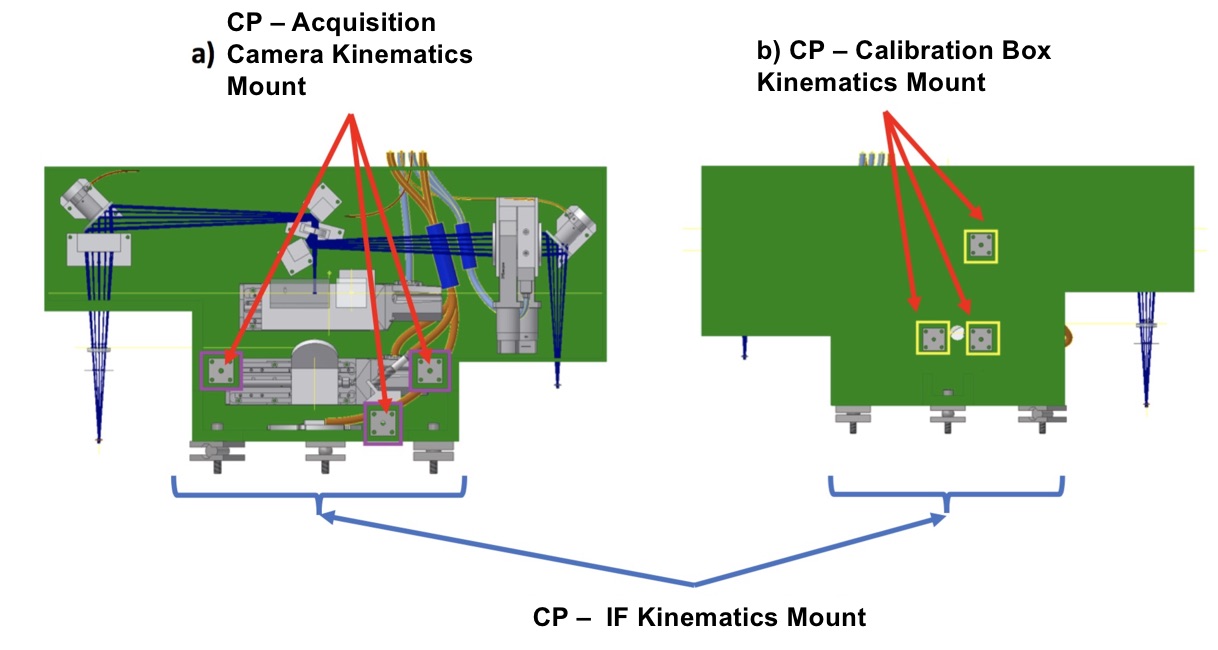}
   \end{tabular}
   \end{center}
   \caption[example] 
%>>>> use \label inside caption to get Fig. number with \ref{}
   { \label{fig:cpic} The mechanical interfaces of common path with the IF, the acquisition camera and the calibration box.}
   \end{figure}

The mechanical interface between the CP and the IF is provided by a large KM connection to be used for the repeatable mounting of the subsystem on the IF flange. 
%The installation of the CP will be done inserting it in the Y direction. 
Two small detachable rails are foreseen to support the CP while moving it toward the KMs.
 Furthermore, the CP provides a small KM connection for the acquisition camera (CAM) subsystem (highlighted in the 3 purple squares in Figure\ \ref{fig:cpic}).  Due to the low weight of the CAM subsystem and its installation in the gravity direction, no guide rails are foreseen. However, 3 temporary soft tipped screws will be installed on it to prevent shocks during the installation of the subsystem. The CP provides a small KM connection for the calibration box (CBX) subsystem (highlighted in the 3 yellow squares in Figure\ \ref{fig:cpic}).
In both interfaces, the KM system is a smaller version of the one used to connect the elements to the IF flange. The part fixed on the CP will consist of 3 steel plates 3 different grooves and a sphere, a cylinder, and a spherical washer.

% Note: If compiling with LaTeX+dvipdf, please ensure images generated from 
% other software packages have their bounding boxes set correctly.

To ensure the maximum stiffness of the CP structure, all the walls are structural with thickness varying between 5 and 20\ mm. The weight of this part is about 22\ kg. Only the cover is not structural and is made of a 3mm thick aluminum plate.
To ensure the maximum repeatability, the KMs are permanently mounted on the structure and characterized with respect to a series of pins on the CP bench. A total of 7 holes are foreseen on the structure: 3 of them (on the bottom) are used for light input and output, 3 (on the 3 other walls) have alignment purpose, 1 on the base to inject the calibration light. The hole for the camera light is placed on the cover.

With reference to Figure\ \ref{fig:soxs2}, the first part of the CP is composed of an instrument shutter, the CBX selector, the CAM selector and a dichroic. The light reflected from the dichroic will go to the UV-VIS spectrograph; in the CP, there will be a flat mirror, an ADC with 2 counter-rotating prisms and a piezo tip-tilt. A small field lens is located above the slit and, even if from the optical point of view is part of the CP, will be located inside the UV-VIS instrument and, therefore, will be presented from the mechanical point of view in the UV-VIS Ref.\ \citenum{rubinetal2018}.
The light transmitted by the dichroic will go to the NIR spectrograph; in the CP, there will be a flat mirror, a piezo tip-tilt and a lens. A small field lens, the pupil stop and the NIR window are located above the slit and, even if from the optical point of view are part of the CP, will be located inside the NIR instrument and, therefore, will be presented from the mechanical point of view in the NIR Ref.\ \citenum{vitalietal2018}.
 
All the elements will be connected with 3 screws and 3 dowel pins. Those elements will be used to guarantee the maximum repeatability in case of mounting and dismounting the elements.
To ease the alignment all the elements will be also mechanical referenced w.r.t the optical path. Each element will also be shimmed between bench/pins and the element stops to have it aligned in all the useful degrees of freedom.

\section{Control Electronics}
\label{subsec:electronics}  % \label{} allows reference to this section

The Instrument Control Electronics (ICE\cite{capassoetal2018}) is based on one main PLC and some I/O modules connected to all subsystems. The modules are connected to the main PLC via the EtherCAT fieldbus.
The PLC offers an OPC-Unified Architecture interface on the LAN. The Instrument Software (INS\cite{riccietal2018}) installed on the Instrument Workstation (IWS) uses this protocol to send commands and read the status to/from the PLC. Electromagnetic compatibility, safety issues, and accessibility for maintenance purpose have driven the whole design. The main function of CP to be controlled are described in Table\ \ref{tab:func}.

\begin{table}[ht]
\caption{Function and motorization of the common path.} 
\label{tab:func}
\begin{center}       
\begin{tabular}{|c|c|c|} 
\hline
\rule[-1ex]{0pt}{3.5ex}  Function& Type & Motorization \\
\hline
\rule[-1ex]{0pt}{3.5ex}  Instrument Shutter & Uni--stable    &   Vincent CS 65 B     \\
\hline
\rule[-1ex]{0pt}{3.5ex}  Calibration Box Selector & Linear stage    & PI L-406.40DD10      \\
\hline
\rule[-1ex]{0pt}{3.5ex}  Acquisition Camera Selector & Linear stage    & PI L-406.40DD10      \\
\hline
\rule[-1ex]{0pt}{3.5ex}  ADC Prism 1 & Rotary stage    &  OWIS DMT100     \\
\hline
\rule[-1ex]{0pt}{3.5ex}  ADC Prism 2 & Rotary stage    &  OWIS DMT100     \\
\hline
\rule[-1ex]{0pt}{3.5ex}  NIR Arm focusing & Linear stage    & PI M-111.1DG      \\
\hline
\rule[-1ex]{0pt}{3.5ex}  Flexure Comp. UV--VIS Arm& Piezo Tip--Tilt    & PI S-330.2SH      \\
\hline
\rule[-1ex]{0pt}{3.5ex}  Flexure Comp. NIR Arm& Piezo Tip--Tilt    &  PI S-330.2SH     \\
\hline
\rule[-1ex]{0pt}{3.5ex}  CP Temperature Sensor& PT--100    &       \\
\hline
\end{tabular}
\end{center}
\end{table}

%\section{Common Path Efficiency}
%\label{sec:efficency}  % \label{} allows reference to this section

\section{Conclusion}
\label{sec:conclusion}  % \label{} allows reference to this section
The SOXS Common Path is the backbone of this instrument connecting the telescope unit (NTT) with all the SOXS  scientific arms (UV-VIS and NIR spectrographs and the acquisition camera) and the scientific arms with the calibration units. We have outlined the main aspects of the SOXS common path, highlighting the optical characteristics, the mechanical interfaces with the other SOXS subsystems and the CP motorized function and service function to be controlled by control electronics and instrument software.
%\acknowledgments % equivalent to \section*{ACKNOWLEDGMENTS}       

% References
\bibliography{report} % bibliography data in report.bib
\bibliographystyle{spiebib} % makes bibtex use spiebib.bst

\end{document}